\documentclass[twocolumn,aps,superscriptaddress,showpacs]{revtex4}

\usepackage{amssymb}
\usepackage{amsmath}
\usepackage{graphicx}
\usepackage[normalem]{ulem}

\begin{document}
\title{Isospin effect on baryon and charge fluctuations from the pNJL model}
\author{He Liu}
\affiliation{Science School, Qingdao University of Technology, Qingdao 266000, China}
\author{Jun Xu\footnote{corresponding author: xujun@zjlab.org.cn}}
\affiliation{The Interdisciplinary Research Center, Shanghai Advanced Research Institute, Chinese Academy of Sciences, Shanghai 201210, China}
\affiliation{Shanghai Institute of Applied Physics, Chinese Academy of Sciences, Shanghai 201800, China}

\begin{abstract}
We have studied the possible isospin corrections on the skewness and kurtosis of net-baryon and net-charge fluctuations in the isospin asymmetric matter formed in relativistic heavy-ion collisions at RHIC-BES energies, based on a 3-flavor Polyakov-looped Nambu-Jona-Lasinio model. With typical scalar-isovector and vector-isovector couplings leading to the splitting of $u$ and $d$ quark chiral phase transition boundaries and critical points, we have observed considerable isospin effects on the susceptibilities, especially those of net-charge fluctuations. Reliable experimental measurements at even lower collision energies are encouraged to confirm the observed isospin effects.
\end{abstract}


\maketitle

\section{Introduction}
\label{introduction}

Studying the hadron-quark phase transition and exploring the phase structure of quantum chromodynamics (QCD) matter are the fundamental goals of relativistic heavy-ion collision experiments. Lattice QCD simulations predict that the transition between the hadronic phase and the partonic phase is a smooth crossover at nearly zero baryon chemical potential ($\mu_B \sim 0$)~\cite{Ber05,Aok06,Baz12a}. At larger $\mu_B$, the transition can be a first-order one based on investigations from theoretical models, see, e.g., studies from the Nambu-Jona-Lasinio (NJL) model and its extensions~\cite{Asa89,Fuk08,Car10,Bra13}. The knowledge of the QCD critical point (CP) in-between the smooth crossover and the first-order phase transition boundaries is important in mapping out the whole QCD phase diagram~\cite{Ste06}. In order to find the signature of the QCD CP at finite $\mu_B$, the Beam Energy Scan (BES) program at the Relativistic Heavy-Ion Collider (RHIC)~\cite{MMA10} has been carrying out $``$low-energy$"$ relativistic heavy-ion collisions and obtained many interesting results. However, heavy-ion collisions with neutron-rich beams produce isospin asymmetric quark matter consisting of different net numbers of $u$ and $d$ quarks, and the isospin degree of freedom is expected to be increasingly important at lower collision energies with larger $\mu_B$, related to the QCD phase structure at finite isospin chemical potentials $\mu_I$~\cite{Son01}. Based on studies from the statistical model~\cite{And09,Sta14,Hat16}, the $\mu_I$ can be as large as 10 MeV at the chemical freeze-out stage of heavy-ion collisions at RHIC-BES energies. The isospin effect is thus expected to influence the search of QCD CP signals in heavy-ion collision experiments.

One of the characteristic feature of the CP is the divergence of fluctuations and the correlation length. Experimentally, such fluctuations in the grand canonical ensemble can be measured from event-by-event observables, such as particle multiplicities, net baryon numbers, or net charge numbers, etc. The singular contribution to the quadratic variance of these observables is related to the correlation length~\cite{Ste09}, which diverges at the CP in the ideal system but mostly limited by the finite size or the finite life time of the system. Investigating the quark matter formed in relativistic heavy-ion collisions, the susceptibilities of conserved quantities, which have their corresponding chemical potentials, carry information of the QCD phase boundary as well as the position of the CP~\cite{Asa09}. The above findings have stimulated experimental measurements of non-Gaussian net-baryon, net-charge, and net-strangeness fluctuations, characterized by the kurtosis and the skewness of their event-by-event distributions, from $\sqrt{s_{NN}}=7.7$ to $200$ GeV at RHIC~\cite{Agg10,Ada14,Ada14c,Ada16,Ada18,Ada19,Ada20a,Ada20b}, and theoretical studies based on lattice QCD calculations~\cite{Baz12,Bor13} as well. For reviews on this topic, we refer the reader to Refs.~\cite{Asa16,Luo17}. However, the isospin effects on these fluctuations have not been seriously investigated. Although lattice QCD can explore the phase transition at finite $\mu_I$~\cite{Kog02,Det12,Bra16}, it suffers from the fermion sign problem at finite $\mu_B$~\cite{Bar86,Kar02,Mur03}. It is of great interest to investigate such isospin effects with an effective QCD model.

In the present study, we investigate the isospin effects on the net-baryon and net-charge fluctuations based on the 3-flavor Polyakov-looped NJL (pNJL) model with isovector couplings in both the scalar and vector channels~\cite{Liu16}, i.e., the scalar-isovector and vector-isovector couplings. The isovector couplings lead to different potentials of $u$ and $d$ quarks in isospin asymmetric quark matter, and server as a possible explanation of the elliptic flow splitting between $\pi^+$ and $\pi^-$ in relativistic heavy-ion collisions~\cite{Liu19}. In addition, the isovector couplings affect the equation of state of isospin asymmetric quark matter and the properties of strange quark stars~\cite{Liu20} as well as hybrid stars~\cite{Liu16}. More importantly, the isovector couplings also lead to the isospin splittings of chiral phase transition boundaries and critical points for $u$ and $d$ quarks in isospin asymmetric quark matter~\cite{Fra03,Tou03,Zha14,Liu16}. It is expected that such isospin splittings may affect the susceptibilities of conserved quantities mentioned above.

\section{PNJL model with isovector couplings}

The thermodynamic potential of the 3-flavor pNJL model with the scalar-isovector and vector-isovector couplings at temperature $T$ can be expressed as~\cite{Liu16}
\begin{eqnarray}\label{eq1}
\Omega_{\textrm{pNJL}} &=& \mathcal{U}(\Phi,\bar{\Phi},T)-2N_c\sum_{i=u,d,s}\int_0^\Lambda\frac{d^3p}{(2\pi)^3}E_i
\notag\\
&-&2T\sum_{i=u,d,s}\int\frac{d^3p}{(2\pi)^3}[\ln(1+e^{-3\beta(E_i-\tilde{\mu}_i)}
\notag\\
&+&3\Phi e^{-\beta(E_i-\tilde{\mu}_i)}
+3\bar{\Phi}e^{-2\beta(E_i-\tilde{\mu}_i)})
\notag\\
&+&\ln(1+e^{-3\beta(E_i+\tilde{\mu}_i)}
+3\bar{\Phi} e^{-\beta(E_i+\tilde{\mu}_i)}
\notag\\
&+&3\Phi e^{-2\beta(E_i+\tilde{\mu}_i)})]
+G_S(\sigma_u^2+\sigma_d^2+\sigma_s^2)
\notag\\
&-&4K\sigma_u\sigma_d\sigma_s
+G_V(\rho_u^2+\rho_d^2+\rho_s^2)
\notag\\
&+&G_{IS}(\sigma_u-\sigma_d)^2
+G_{IV}(\rho_u-\rho_d)^2.
\end{eqnarray}
In the above, the temperature-dependent effective potential $\mathcal{U}(\Phi, \bar{\Phi}, T)$ as a function of the Polyakov loop $\Phi$ and $\bar{\Phi}$ is expressed as~\cite{Fuk08}
\begin{eqnarray}
\mathcal{U}(\Phi,\bar{\Phi},T) &=& -b \cdot T\{54e^{-a/T}\Phi\bar{\Phi}
+\ln[1-6\Phi\bar{\Phi}
\notag\\
&-&3(\Phi\bar{\Phi})^2+4(\Phi^3+\bar{\Phi}^3)]\},
\end{eqnarray}
with $a = 664$ MeV and $b$ = 0.015$\Lambda^3$~\cite{Fuk08}, where $\Lambda = 750$ MeV is the cutoff value in the momentum integral of the second term in Eq.~(\ref{eq1}). The factor $2N_c$ with $N_c=3$ represents the spin and color degeneracy, and $\beta = 1/T$ represents the temperature. $G_S$ and $G_V$ are respectively the scalar-isoscalar and vector-isoscalar coupling constants, $K$ is the coupling constant of the six-point Kobayashi-Maskawa-t'Hooft (KMT) interaction,  and $G_{IS}$ and $G_{IV}$ are respectively the strength of the scalar-isovector and vector-isovector couplings that break the SU(3) symmetry while keeping the isospin symmetry. For the ease of discussions, we define $R_{IS} = G_{IS}/G_S$ and $R_{IV} = G_{IV}/G_S$ as the reduced strength of the scalar-isovector and vector-isovector couplings. The energy $E_i$ of quarks with flavor $i$ is expressed as $E_i(p)=\sqrt{p^2 +M_i^2}$, where $M_i$ is the constituent quark mass. In the mean-field approximation, quarks can be considered as quasiparticles with constituent masses $M_i$ determined by the gap equation
\begin{eqnarray}\label{m}
M_i &=& m_i-2G_S\sigma_i+2K\sigma_j\sigma_k-2G_{IS}\tau_{3i}(\sigma_u-\sigma_d),
\end{eqnarray}
where $m_i$ is the current quark mass, $\sigma_i$ stands for the quark condensate, ($i$, $j$, $k$) is any permutation of ($u$, $d$, $s$), and $\tau_{3i}$ is the isospin quantum number of quark flavor $i$, i.e., $\tau_{3u} = 1$, $\tau_{3d} = -1$, and $\tau_{3s} = 0$. As shown in Eq.~\eqref{m}, $\sigma_d$ and $\sigma_s$ contribute to the constituent quark mass of $u$ quarks as a result of the six-point interaction and the scalar-isovector coupling. Similarly, the effective chemical potential expressed as
\begin{eqnarray}
\tilde{\mu}_i&=&\mu_i+2G_V\rho_i+2G_{IV}\tau_{3i}(\rho_u-\rho_d)
\end{eqnarray}
has also the contribution from quarks of other isospin states through the vector-isovector coupling. The net quark number density of flavor $i$ can be calculated from
\begin{eqnarray}
\rho_i=2N_c\int(f_i-\bar{f_i})\frac{d^3p}{(2\pi)^3},
\end{eqnarray}
where
\begin{eqnarray}
f_i=\frac{1+2\bar{\Phi}\xi_i+\Phi\xi_i^2}{1+3\bar{\Phi}\xi_i+3\Phi\xi_i^2+\xi_i^3}
\end{eqnarray}
and
\begin{eqnarray}
\bar{f_i}=\frac{1+2\Phi{\xi}'_i+\bar{\Phi}{{\xi}'_i}^2}{1+3\Phi{\xi}'_i+3\bar{\Phi}{{\xi}'_i}^2+{{\xi}'_i}^3}
\end{eqnarray}
are the effective phase-space distribution functions for quarks and antiquarks in the pNJL model, with $\xi_i= e^{(E_i-\tilde{\mu_i})/T}$ and ${\xi}'_i= e^{(E_i+\tilde{\mu_i})/T}$.

In the present study, we adopt the values of parameters~\cite{Bra13, Lut92} as $m_u = m_d = 3.6$ MeV, $m_s = 87$ MeV, $G_S\Lambda^2 = 3.6$, and $K\Lambda^5 = 8.9$. Although the lattice QCD calculations of the susceptibility favor $G_V=0$~\cite{Kun91,Ste11}, a finite $G_V$ leads to different dynamics of quarks and antiquarks and helps to explain the elliptic flow splittings between protons and antiprotons in RHIC-BES experiments~\cite{Xu14}. On the other hand, the effect of $G_V$ on the position of the critical point and the susceptibility is well known~\cite{Asa89,Fuk08,Car10,Bra13}. Since the purpose is not to study the effect of $G_V$ on the structure of the phase diagram, it is set to 0 in the present study. The numerical calculation is based on the following equations
\begin{eqnarray}
\frac{\partial\Omega_{\textrm{pNJL}}}{\partial\sigma_u}
=\frac{\partial\Omega_{\textrm{pNJL}}}{\partial\sigma_d}
=\frac{\partial\Omega_{\textrm{pNJL}}}{\partial\sigma_s}
=\frac{\partial\Omega_{\textrm{pNJL}}}{\partial\Phi}
=\frac{\partial\Omega_{\textrm{pNJL}}}{\partial\bar{\Phi}}
=0,
\notag\\
\end{eqnarray}
in order to obtain the values of $\sigma_u$, $\sigma_d$, $\sigma_s$, $\Phi$, and $\bar{\Phi}$ at the minimum thermodynamic potential in the pNJL model.

\section{Isospin properties of higher-order susceptibilities}

\begin{figure*}[ht]
\includegraphics[angle=0,scale=0.65]{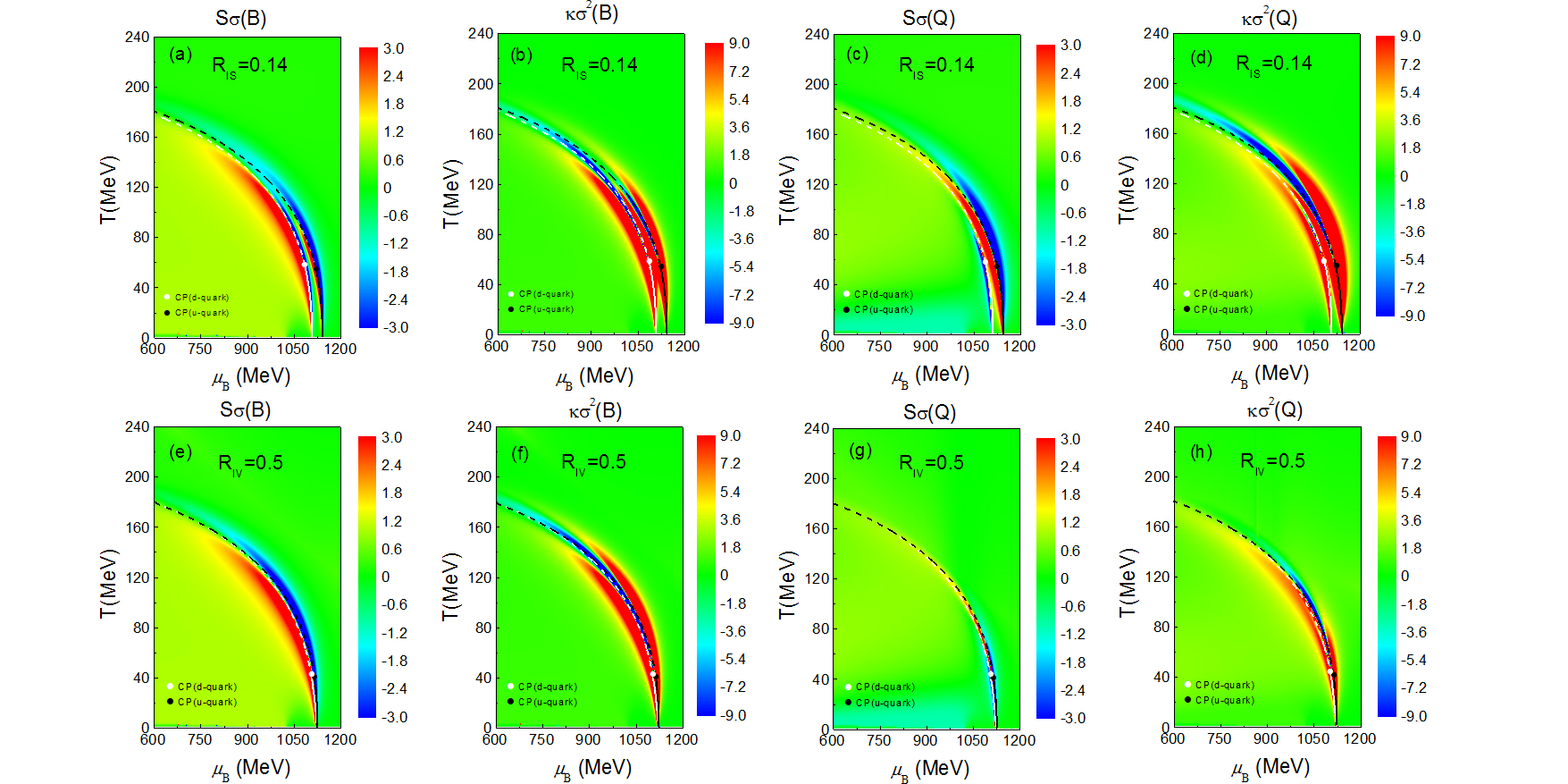}
\caption{(color online) Skewness and kurtosis of net-baryon (B) (left) and net-charge (Q) (right) fluctuations in the $\mu_B - T$ plane with finite scalar-isovector (upper panels) and vector-isovector (lower panels) coupling constants and the empirical relation $\mu_I = -0.293 - 0.0264\mu_B$ with both $\mu_B$ and $\mu_I$ in MeV.  The chiral phase transition boundaries and the corresponding critical points (CP) of $u$ (black) and $d$ (white) quarks are plotted in all panels for reference. } \label{phase}
\end{figure*}
In the following, we consider fluctuation moments of conserved quantities, such as net-baryon and net-charge fluctuations, from the above 3-flavor pNJL model. The $n$th-order susceptibility representing the cumulant of a given conserved quantity in the grand ensemble can be expressed as the derivative of the thermodynamic potential as~\cite{Asa16}
\begin{eqnarray}
\chi^{(n)}_X = \frac{\partial^n(-\Omega/T)}{\partial(\mu_X/T)^n},
\end{eqnarray}
where $\mu_X$ represents the baryon ($\mu_B$) or the charge ($\mu_Q$) chemical potential. Numerically, the isospin chemical potential $\mu_I$ and the charge chemical potential $\mu_Q$ are equal to each other~\cite{Baz12,Che15}. In the present study, we use the empirical relation between the isospin chemical potential and the baryon chemical potential, i.e., $\mu_I = -0.293 - 0.0264\mu_B $ with both $\mu_B$ and $\mu_I$ in MeV, determined from the statistical model fits of the particle yield in Au+Au collisions at center-of-mass energy $\sqrt{s_{NN}}$ from 7.7 GeV to 200 GeV~\cite{And09,Sta14,Hat16}. For the strangeness chemical potential, we take the empirical relation as $\mu_S = 1.032 + 0.232\mu_B $~\cite{And09,Sta14,Hat16}. Note that the chemical potentials of $u$, $d$, and $s$ quarks in the 3-flavor pNJL model can be expressed in terms of $\mu_B$, $\mu_I$, and $\mu_S$. The relations between these chemical potentials give some constraints on the quark and antiquark number densities in the phase diagram, and they are used to illustrate qualitatively the isospin effect from the isovector interactions. The higher-order susceptibilities are related to the skewness $S$ and kurtosis $\kappa$ measured experimentally in relativistic heavy-ion collisions through the relations~\cite{Asa16}
\begin{eqnarray}
S\sigma = \frac{\chi^{(3)}}{\chi^{(2)}},
\quad  \kappa\sigma^2 = \frac{\chi^{(4)}}{\chi^{(2)}},
\end{eqnarray}
where $\sigma$ is the variance of the conserved quantity. The subscript of the net baryon (B) or the net charge (Q) is omitted in the above equations. We note that from Eq.~(\ref{eq1}) the Polyakov-loop interaction only contributes to the single quark fluctuations, while the fluctuations of the flavor mixed states are from the KMT interaction, the scalar-isovector interaction, and the vector-isovector interaction. Although the pNJL model has the deficiency of lacking of hadronic effects, in the present study we are interested in the region close to the chiral phase transition boundary as well as the critical point. It is noteworthy that the pNJL model has been used to study the phase diagram and the susceptibility of net-baryon and net-charge fluctuations in many works~\cite{Fuk08,Car10,Rat07,Saw07,Kas08,Bub09,Cri10,Bha10,Bha15,Bha17,Sha18,Fer18,Li19}. Since the hadronic effects become dominate at lower beam energies or temperatures, there are also studies on the susceptibilities based on pure hadronic models or both hadronic models and quark models, see, e.g., Refs.~\cite{Fuk15,Nah15,Mis16,Muk17,Vov17,Ada17a,Fu17}.

We begin our discussion on the higher-order net-baryon and net-charge susceptibilities from the 3-flavor pNJL model in the $\mu_B - T$ plane with various isovector coupling constants, as shown in Fig.~\ref{phase}. As shown in Fig.~3 of Ref.~\cite{Liu16}, $R_{IS}=0.14$ and $R_{IV}=0.5$ lead to the splittings of $u$ and $d$ quark chiral phase transition boundaries as well as their critical points, which are plotted in all panels of Fig.~\ref{phase} for references. These chiral phase transition boundaries indicate where there are dramatic changes of the quark mass or the quark condensate in Eq.~(\ref{m}). As also discussed in Ref.~\cite{Liu16}, the splitting of the chiral phase transition related to the effective quark mass splitting is mostly sensitive to $R_{IS}$, while the $R_{IV}$ term has only the secondary effect. We note that generally the relativistic heavy-ion collision system goes through a trajectory in the 3-dimensional phase diagram ($T$, $\mu_B$, $\mu_I$). The empirical relation $\mu_I = -0.293 - 0.0264\mu_B$ extracted from chemical freeze-out is only well valid near the phase boundary, and this is exactly the region where we are interested. As seen in Fig.~\ref{phase}, chiral phase transition boundaries separate the red and blue areas for skewness results, representing respectively the positive and negative values of $S\sigma$. For kurtosis results, however, the chiral phase transition boundaries go through the blue areas, and the critical points stand at the ends of the blue areas. It is also interesting to see that the skewness of net-charge fluctuations gives the different orders of the red and blue areas compared to that of net-baryon fluctuations.

\begin{figure*}[ht]
\includegraphics[scale=0.5]{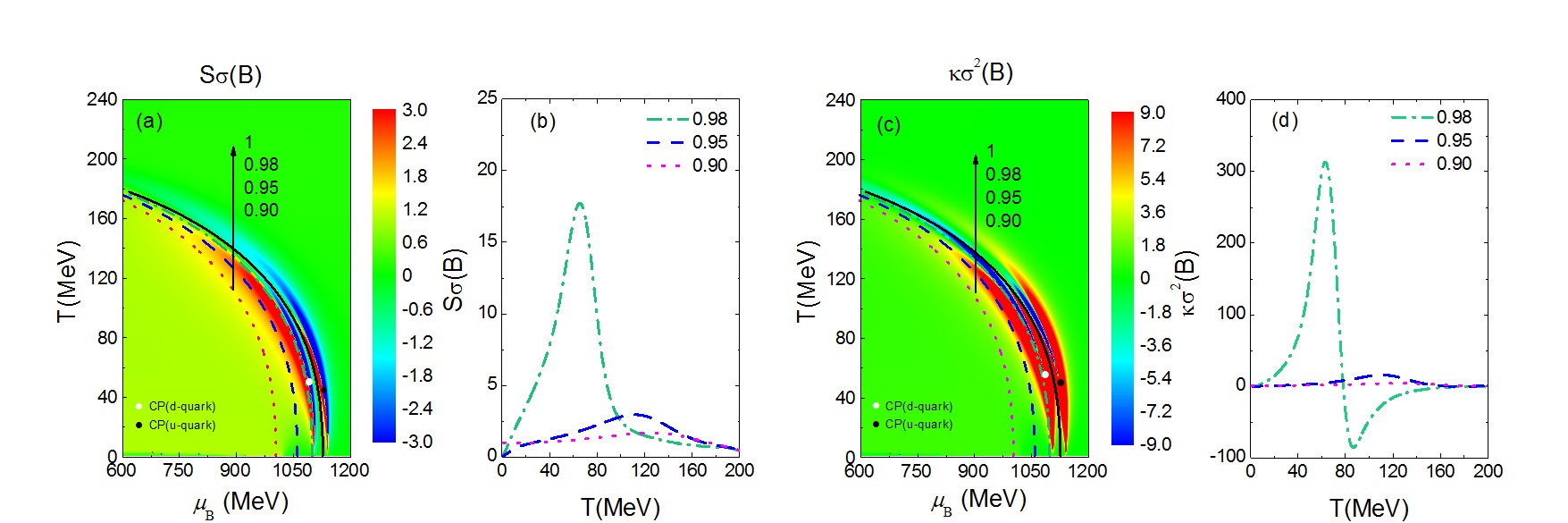}
\includegraphics[scale=0.51]{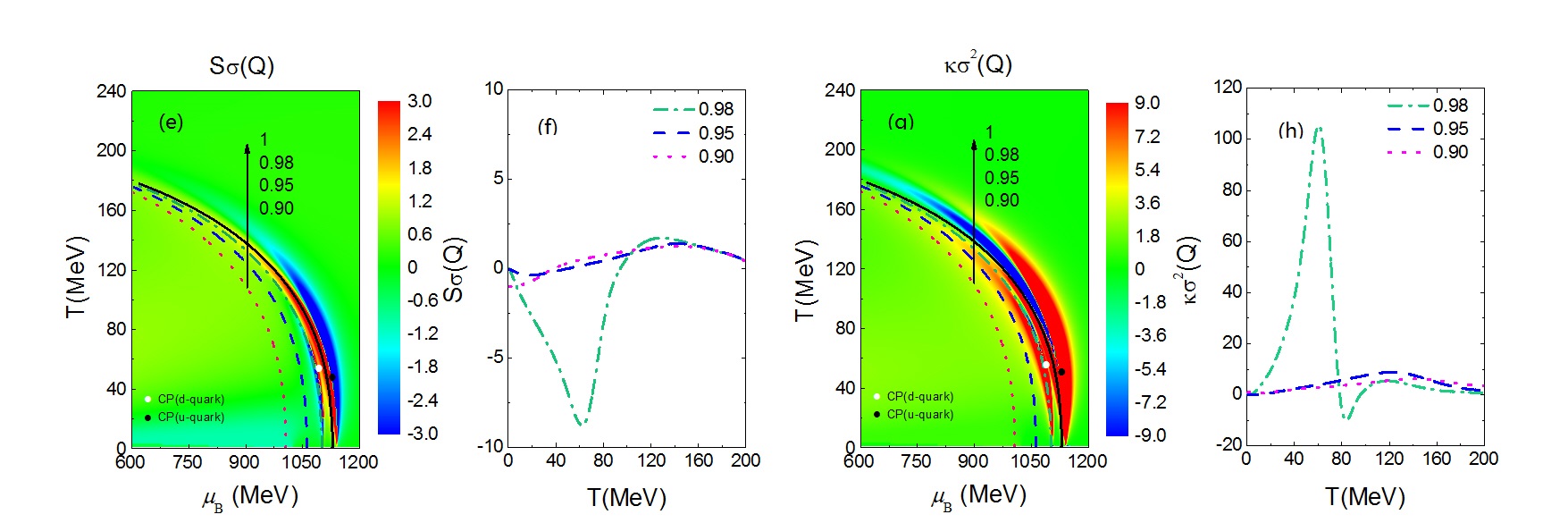}
\caption{(color online) Density plot of $S\sigma$ (left) and $\kappa\sigma^2$ (right) in the $\mu_B - T$ plane as well as those along the different hypothetical chemical freeze-out lines with the scalar-isovector coupling constant $R_{IS}=0.14$ and the empirical relation $\mu_I = -0.293 - 0.0264\mu_B$ with both $\mu_B$ and $\mu_I$ in MeV. The solid line represents the averaged chiral phase transition boundary for $u$ and $d$ quarks, while the other lines are hypothetical chemical freeze-out lines by rescaling $\mu_B$ of the solid line with factors of $0.98$, $0.95$, and $0.90$. The upper panels are the net-baryon (B) susceptibilities while the lower panels are the net-charge (Q) susceptibilities.} \label{phase-boundary}
\end{figure*}

In relativistic heavy-ion collision experiments, the net-baryon and net-charge fluctuations are measured at the chemical freeze-out. It is well-known that the pseudocritical temperature from lattice QCD calculations at zero baryon density is the same as that extracted from the statistical model at top RHIC energy or LHC energy, leading to the conclusion that the chemical freeze-out happens right after the hadron-quark phase transition in ultra-relativistic heavy-ion collisions. In heavy-ion collisions at RHIC-BES energies, when the chemical freeze-out happens is not well determined. There are several empirical criteria for the chemical freeze-out in relativistic heavy-ion collisions, such as fixed energy per particle at about 1 GeV, fixed total density of baryons and antibaryons, fixed entropy density over $T^3$, as well as the percolation model and so on (see Ref.~\cite{Cle06} and references therein). In order to compare qualitatively the higher-order susceptibilities from the pNJL model with experimental results, we obtain the hypothetical chemical freeze-out lines by rescaling $\mu_B$ of the averaged chiral phase transition boundaries of $u$ and $d$ quarks with factors of $0.98$, $0.95$, and $0.90$, corresponding respectively to the dash-dotted, dashed, and dotted curves in Fig.~\ref{phase-boundary}. We note that a similar assumption of the chemical freeze-out lines was made in Ref.~\cite{Che17}, and the present hypothetical chemical freeze-out lines are always below the chiral phase transition boundaries of both $u$ and $d$ quarks. For the net-baryon susceptibility shown in the upper panels of Fig.~\ref{phase-boundary}, it is seen that $S\sigma(B)$ has one positive peak while $\kappa\sigma^2(B)$ has a positive and a negative peak along the chemical freeze-out lines, and the peaks are sharper if the hypothetical chemical freeze-out lines are closer to the chiral phase transition boundary, as also shown in many previous works. The critical point of the $d$ quark chiral phase transition is always at the low-temperature or low-energy side of the positive peaks for both $S\sigma(B)$ and $\kappa\sigma^2(B)$, and the distance between the positive peaks and the critical point is related to that between the chemical freeze-out line and the chiral phase transition boundary. For the net-charge susceptibility shown in the lower panels of Fig.~\ref{phase-boundary}, we have also observed sharper peaks if the hypothetical chemical freeze-out line is closer to the chiral phase transition boundary. It is seen that $S\sigma(Q)$ has a negative peak at lower temperatures/energies and a positive peak at higher temperatures/energies, and the broad positive peak of $\kappa\sigma^2(Q)$ turns to two positive peaks and a negative one if the hypothetical chemical freeze-out line is very close to the chiral phase transition boundary. Again, the critical point of the $d$ quark chiral phase transition is at the low-temperature or low-energy side of the negative peak for $S\sigma(Q)$ or the positive peak for $\kappa\sigma^2(Q)$.

 \begin{figure*}[ht]
\includegraphics[scale=0.5]{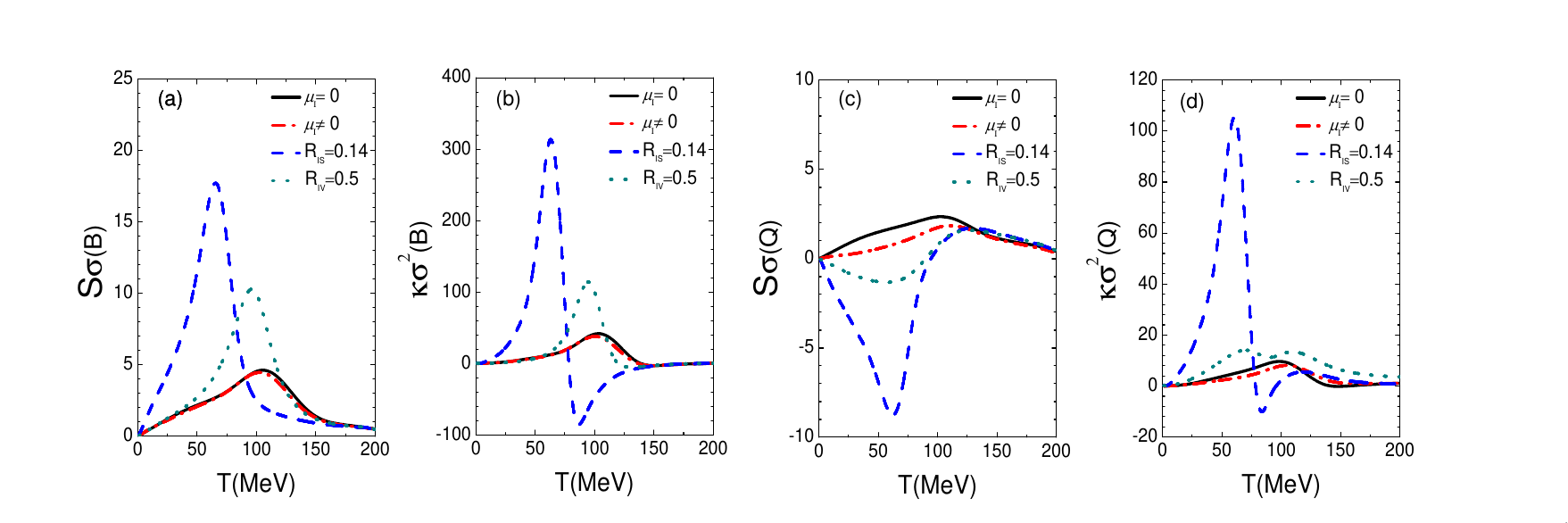}
\caption{(color online) $S\sigma$ and $\kappa\sigma^2$ for net-baryon (B) and net-charge (Q) fluctuations along the closest hypothetical chemical freeze-out line to the phase boundary with four representative scenarios of isospin chemical potentials and isovector couplings.}\label{boundary}
\end{figure*}

In order to compare the higher-order susceptibilities with and without the isospin effect, we display in Fig.~\ref{boundary} the skewness and kurtosis of net-baryon and net-charge fluctuations with four representative scenarios of isospin chemical potentials and isovector couplings: $\mu_I = 0$, $R_{IS} = 0$, $R_{IV} = 0$; $\mu_I = -0.293 - 0.0264\mu_B$, $R_{IS} = 0$, $R_{IV} = 0$; $\mu_I = -0.293 - 0.0264\mu_B$, $R_{IS} = 0.14$, $R_{IV} = 0$; $\mu_I = -0.293 - 0.0264\mu_B$, $R_{IS} = 0$, $R_{IV} = 0.5$, with both $\mu_I$ and $\mu_B$ in MeV. In order to illustrate the largest possible isospin effect, the susceptibilities are calculated along the closest hypothetical chemical freeze-out line to the phase boundary, i.e., using the rescaling factor of 0.98. It is seen that even without isovector couplings, the skewness and kurtosis can be slightly different for $\mu_I=0$ and $\mu_I \neq 0$, especially for net-charge susceptibilities. With finite isovector coupling constants, the isospin effect is largely enhanced. The peaks of the net-baryon susceptibilities move to the low-temperature or low-energy side, especially for $R_{IS}=0.14$. The general shape of the net-baryon susceptibility is qualitatively consistent with the experimental data~\cite{Ada14,Luo}. The net-baryon susceptibility is not a unique probe of the isospin effect, since it is largely affected by other effects as well, such as the vector-isoscalar coupling ($G_V$ term in Eq.~(\ref{eq1}))~\cite{Fuk08,Car10}. On the other hand, the isospin couplings affect dramatically the net-charge susceptibilities. It is seen that the isovector couplings lead to a negative peak at lower temperatures/energies for $S\sigma(Q)$. For $\kappa\sigma^2(Q)$, the isovector couplings lead to two peaks and $R_{IS}=0.14$ is the only scenario that leads to negative values. The experimental results from STAR and PHENIX Collaborations for net-charge susceptibilities are not consistent with each other yet~\cite{Ada14c,Ada16}. So far the experimental results for $S\sigma(Q)$ seem to be positive above $\sqrt{s_{NN}}=7.7$ GeV from both STAR and PHENIX measurements, and it is of great interest to distinguish different scenarios if reliable measurements are done at even lower collision energies. For the $\kappa\sigma^2(Q)$ results, the STAR results lead to negative values at lower $\sqrt{s_{NN}}$~\cite{Ada14c} while the PHENIX results remain positive at all energies~\cite{Ada16}. It is again of great interest to check experimentally whether another peak appears at even lower collision energies, as seen from the $\kappa\sigma^2(Q)$ results with $R_{IS}=0.14$.

\begin{figure*}[ht]
\includegraphics[scale=0.5]{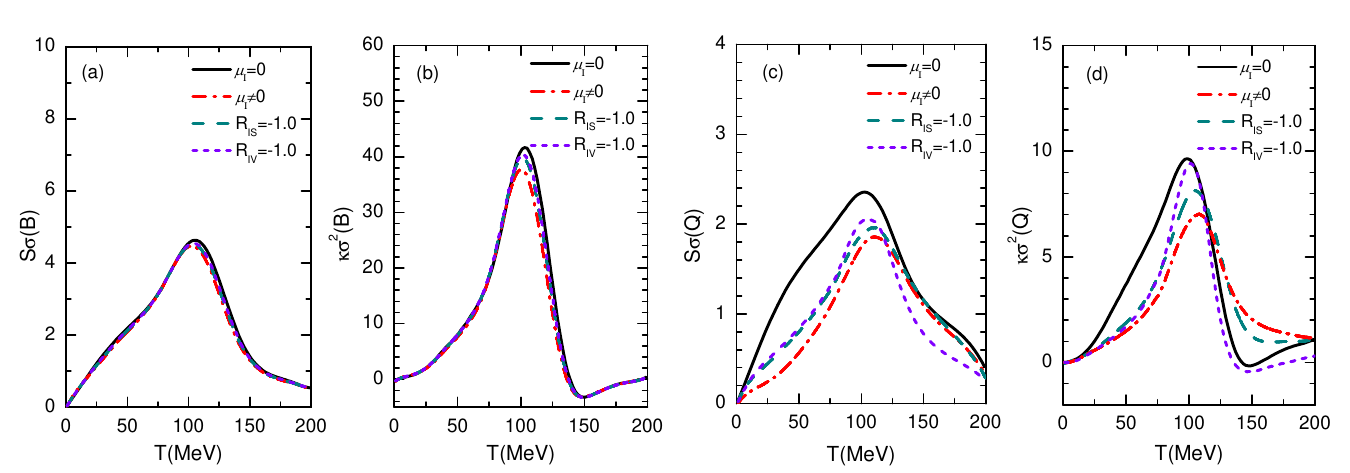}
\caption{(color online) Similar to Fig.~\ref{boundary} but results from $R_{IS}=-1$ and $R_{IV}=-1$ are compared with others.}\label{fig4}
\end{figure*}

The scalar-isovector and vector-isovector coupling constants $G_{IS}$ and $G_{IV}$ are considered as free parameters. In the above study we mainly focused on the extreme case of $R_{IS}=0.14$ and $R_{IV}=0.5$, which lead to the maximum splitting of $u$ and $d$ quark chiral phase transition boundaries as well as their critical points. As shown in Ref.~\cite{Liu16}, for even larger isovector coupling constants, the $d$ quark chemical potential near the phase boundary is comparable to the cutoff value $\Lambda$ in the momentum integral. For completeness, we have also compared results from negative values of isovector coupling constants, i.e., $\mu_I = -0.293 - 0.0264\mu_B$, $R_{IS}=-1$, $R_{IV}=0$; $\mu_I = -0.293 - 0.0264\mu_B$, $R_{IS}=0$, $R_{IV}=-1$ with others in Fig.~\ref{fig4}. The negative isovector couplings have negligible effects on the phase diagram at finite isospin chemical potentials as shown in Ref.~\cite{Liu16}, and they have respectively negligible and small effects on the net-baryon and net-charge susceptibilities as shown in Fig.~\ref{fig4}.

Although the present study relies on an ideal thermodynamic calculation, it gives the intuitive picture of the isospin effect. To take into account the detailed hadronic effects on the susceptibilities, such as hadronic rescatterings, resonance decays, and the global conservation law, etc., one may need to employ transport models and use the same cuts as in the experimental analysis. Such study goes beyond the present scope.

\section{Summary and outlook}

Based on the 3-flavor Polyakov-looped Nambu-Jona-Lasinio model with the scalar-isovector and vector-isovector couplings, we have studied the higher-order susceptibilities of net-baryon and net-charge fluctuations in isospin asymmetric matter formed in relativistic heavy-ion collisions at RHIC-BES energies. Despite a few assumptions made in the study, the isospin effect on the susceptibilities as a result of the isovector interactions is robust. Although negative isovector couplings will not affect much the phase diagram and the susceptibilities, positive values of the isovector coupling constants seperate the $u$ and $d$ quark chiral phase transition boundaries as well as their critical points at finite isospin chemical potentials. This moves the peaks of the net-baryon susceptibilities to the low-temperature or low-energy side, and largely changes the shape of the net-charge susceptibilities, i.e., an additional negative peak appears in the skewness results and two positive peaks appear in the kurtosis results along the hypothetical chemical freeze-out line, if it is very close to the chiral phase transition boundary.

The peak of the susceptibility reveals the position of the critical point, which is affected not only by the isovector couplings but by the vector-isoscalar coupling as well. Additional constraints from observables are needed to extract values of these coupling constants from the theoretical side. In order to have a more direct comparison with the experimental data, one has to employ transport models~\cite{Xu14} that take into account detailed hadronic effects and apply the same cuts as in the experimental analysis. A recent lattice QCD calculation using Taylor expansion of the chemical potentials disfavors the critical point at $\mu_B/T<2$ and $T>135$ MeV~\cite{LQCD17}. On the experimental side, the peak at lower temperatures/energies is not observed or confirmed yet. It is of great interest to confirm our findings by measuring the net-charge susceptibility at even lower collision energies. Efforts from both the theoretical and experimental sides may be helpful in extracting the strength of the isovector couplings or even the information of the isospin dependence of the QCD phase diagram.

\begin{acknowledgments}
This work was supported by the National Natural Science Foundation of China under Grant No. 11922514.
\end{acknowledgments}

\end{document}